\documentclass[epj,final]{svjour}
\usepackage{graphicx}
\def\be{\begin{eqnarray}}
\def\ee{\end{eqnarray}}
\def\bc{\begin{center}}
\def\ec{\end{center}}

\def\rmd{{\rm d}}
\def\om{\omega}

\newcommand{\lsim}{\stackrel{\scriptstyle <}{\phantom{}_{\sim}}}
\newcommand{\gsim}{\stackrel{\scriptstyle >}{\phantom{}_{\sim}}}

\begin{document}
\title{Mechanism of $r$-mode stability in young rapidly rotating pulsars}
\author{E. E. Kolomeitsev\inst{1} \and D. N. Voskresensky\inst{2}}
\institute{Matej Bel  University, SK-97401 Banska Bystrica, Slovakia
\and
National Research Nuclear
University (MEPhI), 115409 Moscow, Russia}
%=================================================================
\abstract{ We demonstrate that stability of $r$-modes in young
rapidly rotating pulsars might be explained if one takes into
account strong medium modifications of the nucleon-nucleon
interaction because of the softening of pionic degrees of freedom
in dense nucleon matter.  Presence of the efficient direct Urca
processes is not required. Within our model the most rapidly
rotating observed young pulsar PSR J0537-6910 should  have the
mass $\geq 1.8M_{\odot}$.}
%\date{\today}
%\keywords{Fermi liquid, superfluidity, Cooper pair breaking, neutrino emission, vertex function correction }
%\keywords{}
\maketitle
%%%%%%%%%%%%%%%%%%%%%%%%%%%%%%%%%%%%%%%%%%%%%%%%%%%%%%%%%%%%%%%%%%
%\noindent
%  PSR J2022+3842 with $P=48.6$~ms, $\nu=20$~Hz, age: 8.9 kyr\\
%  PSR J0537-6910 with $P=16$~ms, $\nu=62$~Hz age: 5kyr\\
%  PSR B0531+21  with $P=33$~ms, $\nu=30$~Hz age: 1.24kyr

Within the standard scenario of the neutron star (NS) formation in the
supernova explosion~\cite{populsint} the NS is typically formed
rapidly rotating with an initial rotation frequency $\nu_{\rm
in}\sim 1$kHz. However the majority of young pulsars ($<10^5$~yr of age) have
rotation frequencies less than 10~Hz, and the fastest young pulsar observed so far is PSR
J0537-6910~\cite{Marshall} with  $\nu_{\rm max}^{\rm young}=62$~Hz
and age 5~kyr.
%%%%%%%%%%%%%%%%%%%%%%%%%%%%%%%%%%%%%%%%%
An efficient mechanism allowing to decelerate the rotation of a
neutron star already at an early stage of its evolution concerns
the $r$-mode instability predicted in~\cite{Andersson:1997xt}, see
also review~\cite{Anders-Kokkotas00}. The $r$-mode oscillations
lead to emission of gravitational waves, which carry away most
part of the initial angular momentum of a star and the star
rotation decelerates rapidly. The $r$-mode amplitude would grow
exponentially for any rotation frequency, if not a damping because
of a viscosity of warm neutron star matter~\cite{Lindblom:1998wf}.
The typical temperature in the interior of a pulsar of age $\sim
5$~kyr is $\sim 2\cdot 10^7\mbox{--} 5\cdot 10^8$~K, depending on
the  mass of the star and the cooling scenario,
see~\cite{Blaschke:2013vma}. However, the $r$-mode instability
proves to be strongest at higher temperatures $T\sim 10^{9}$~K.
Hence, young pulsars have necessarily passed through an
instability phase during their early history.
%%%%%%%%%%%%%%%%%%%%%%%%%%%%%%%%%%
Thus, to verify the $r$-modes instability scenario for the
star-rotation braking one should demonstrate that after the
instability phase the pulsar frequencies remain larger than
$\nu_{\rm max}^{\rm young}$.
%%%%%%%%%%%%%%%%%%%%%%%%%%%%%%%%%%
Different mechanisms for suppression of the instability were studied.
Most attempts were spent to find appropriate arguments to increase the values of shear and bulk viscosities. Finally, it was concluded, e.g. see~\cite{Anders-Kokkotas00}, that the minimum value of the frequency at the $r$-mode stability boundary $\nu_c
(T)$ is smaller than the value $\nu_{\rm max}^{\rm young}$, if one
uses standard dissipation mechanisms only. The problem proved to be even more serious after Ref.~\cite{Shternin:2008es} demonstrated that the lepton shear viscosity should be strongly suppressed by in-medium polarization effects, compared to the result of the previous calculation~\cite{FlowersItoh}.\\
\indent In this letter we investigate the $r$-mode stability of
young pulsars taking into account polarization effects in nuclear
matter, among which the most efficient is the softening of pionic
degrees of freedom with an increase of the nucleon density. This
effect proves to be important in description of many phenomena in
atomic nuclei, neutron stars, and heavy-ion
collisions~\cite{Migdal:1990vm}. In particular, incorporating the
above effects, the ``nuclear medium cooling'' scenario developed
in~\cite{Blaschke:2013vma,Voskre-cool,BGV} allows to fit all
existing data on the pulsar surface temperatures, including the
recent data on the cooling of the young pulsar in Cassiopea~A. We
will incorporate mentioned effects  in the $r$-mode dissipation
mechanisms. Also we include new contributions, such as the
neutrino shear viscosity in the neutrino trapping region and
radiative bulk viscosity calculated with account for nucleon
medium polarization effects.\\
\indent The characteristic time scale of the $r$- mode amplitude evolution is given by
%%%%%%%%%%%%%%%%%%%%%%%%%%%%%%%%
$
\tau^{-1}=\tau^{-1}_{\eta}+\tau^{-1}_{\zeta}-\tau^{-1}_G \,,
$
%%%%%%%%%%%%%%%%%%%%%%%%%%%%%%%%
where $\tau_G$ is the typical time of the gravitational radiation,
$\tau_{\eta}$ denotes the relaxation time induced by the shear
viscosity, $\tau_{\zeta}$ stands for the relaxation time induced
by the bulk viscosity. The $r$-modes are unstable, if
$\tau^{-1}<0$. The gravitational time for the most unstable mode,
which oscillation frequency is related to the angular velocity of
the pulsar ($\Omega =2\pi\nu$) as $\om=4\Omega/3$, is equal
to~\cite{Lindblom:1998wf}
%%%%%%%%%%%%%%%%%%%%%%%%%%%%%%%%%%%%%%%%%%%%%%%%%%%%%
$
\tau_G^{-1}=6.4\cdot 10^{-2}~[{\rm Hz}]\, R_{6}^7\, \Omega_4^6
\, \rho_{\rm cen}/\rho_0\,,
$
%%%%%%%%%%%%%%%%%%%%%%%%%%%%%%%%%%%%%%%%%%%%%%%%%%%%%
where $R_6=R/({\rm 10^6\,cm})$, $\Omega_4=\Omega/({\rm 10^4 Hz})$
and $\rho_0=m_N n_0= 2.63\cdot 10^{14}\,{\rm g/cm^3}$ is the mass
density of the  nuclear matter at saturation, with $m_N=938$~MeV
being the nucleon mass in vacuum. The central mass density of the
star $\rho_{\rm cen}$ depends on the neutron star mass. The
damping times of the $r$-modes can be written
as~\cite{Lindblom:1998wf}
%%%%%%%%%%%%%%%%%%%%%%%%%%%%%%%%%
$
\tau_\eta^{-1} = 6.0\cdot 10^{-5}\,[{\rm Hz}]\,
\langle \eta_{20} \rangle_4 \, R_6^{-2} \, \rho_0/\rho_{\rm cen}\,,
$
%%%%%%%%%%%%%%%%%%%%%%%%%%%%%%%%%
and
$%\begin{equation}
\tau_\zeta^{-1} = 2.2\cdot 10^{-7}\,[{\rm Hz}]\, R_6^4\, \Omega^4_4 \langle \zeta_{20}\,[1+0.86 ({r}/{R})^2]\rangle_{8}
(\frac{M_\odot}{M})^2\frac{\rho_0}{\rho_{\rm cen}}\,,
\nonumber
$%\end{equation}
where $\eta_{20}$, $\zeta_{20}$ stand for the shear and bulk
viscosities measured in units of $10^{20}$~g/(cm$\cdot$s). The
angular brackets mean the averaging
$\big\langle\dots\big \rangle_n = {R^{-(n+1)}}
\int_0^R (\dots)\,r^{n} \,\rmd r \,. $

%%%%%%%%%%%%%%%%%%%%%%%%%%%%%%%%%%%%%%%%%%%%%%%%%%%%%%%%%%%%%%%%%
%%%%%%%%%%%%%%%%%%%%%%%%%%%%%%%%%%%%%%%%%%%%%%%%%%%%%%%%%%%%%%%%%
For the nucleon densities $n>0.6 n_0$ we exploit the HDD equation
of state (EoS) constructed in\cite {Blaschke:2013vma} to be close
to the realistic Akmal-Pandharipande-Ravenhall (APR)  A18+$\delta
v+$UIX$^*$ EoS~\cite{APR} for densities $n\lsim 4n_0$. On the
contrary to the APR EoS, the HDD EoS is causal for all densities,
producing the maximum mass $M_{\rm max}\simeq 2.05M_{\odot}$,
being in agreement with observations \cite{maxmass}. As for the
APR EoS, the one-nucleon direct Urca (DU) processes with
electrons, $n\to p+e+\bar{\nu}$, start to contribute only for
densities $n>5n_0$, i.e.,  for stars with masses $M>M_c^{\rm
DU}\simeq 1.9M_{\odot}$. Dependence of the effective nucleon mass
on the nucleon density is parameterized~\cite{Migdal:1990vm} as
$m_N^*\approx m_n^* \approx m_p^* \approx [1-0.15(n/n_0)^{1/2}]\,
m_N$. For the density $n\simeq 0.6\, n_0$ we match the HDD EoS
with the Fried\-man-Pan\-dha\-ri\-pan\-de-Skyrme EoS ~\cite{HP04}
that we use for lower densities. As in all previous papers
discussing $r$-mode instability, we will use a simplifying
assumption of a homogeneous distribution of the temperature in the
star core.

%%%%%%%%%%%%%%%%%%%%%%%%%%%%%%%%%%%%%%%%%%%%%%%%%%%%%%%%%%%%%%%%%%%
%%%%%%%%%%%%%%%%%%%%%%%%%%%%%%%%%%%%%%%%%%%%%%%%%%%%%%%%%%%%%%%%%%%
For calculations of different partial contributions to the
neutrino emissivities and viscosities one often uses the free
one-pion exchange (FOPE) model or, sometimes, the free
nucleon-nucleon ($NN$) cross-section corrected by Pauli blocking
~\cite{Shternin:2008es,Friman:1978zq}. Alternatively we exploit
here the $NN$ interaction constructed within the Fermi-liquid
approach~\cite{M67}, wherein nucleon particle-hole excitations are
taken into account explicitly while other processes are
incorporated as a phenomenologically parameterized residual
interaction. Pionic modes are soft ($m_{\pi}\ll m_N$, $m_\pi$ is
the pion mass). Therefore, they are treated explicitly on equal
footing with the nucleon particle-hole modes. As the result, the
main contribution to the $NN$ interaction at densities $n\gsim
n_0$ is given by the medium one-pion exchange (MOPE), whereas the
relative contribution of the residual interaction diminishes with
increasing density because of polarization effects, see
\cite{Migdal:1990vm} and references therein.

Allowing for nucleon superfluidity in the star interior, for the
neutron and proton $^1$S$_0$ pairing gaps  we use same
parameterizations as in~\cite{Blaschke:2013vma,BGV} (set I shown
in Fig. 2 in \cite{Blaschke:2013vma}). The critical temperature of
the neutron triplet pairing was argued~\cite{Schwenk} to be
strongly reduced to the values $\sim 10^8$~K because of a
medium-induced spin-orbit interaction. This choice has been
exploited in the nuclear medium cooling
scenario~\cite{Blaschke:2013vma,BGV} and we continue to use it
here.

\begin{figure}
\includegraphics[width=8cm]{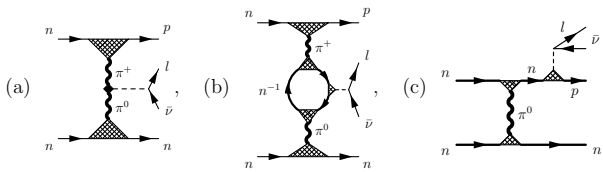}
\caption{Set of diagrams determining the MMU reactions.}
\label{fig:MMU-react}
\end{figure}

For calculations of the $r$-mode damping times $\tau_\eta$ and
$\tau_\zeta$ we need  shear and bulk viscosities. The bulk
viscosity is presented as the sum of three contributions, the
collisional term, the ``soft-mode" term and the radiative term.
The collisional term, determined mainly by $nn$ collisions is
found to be small. The soft-mode contribution to the bulk
viscosity is induced in pulsating medium by reactions generated by
charged weak currents. The radiative bulk viscosity contributions
prove to be of the same order as the soft-mode ones.

Within the ``minimal cooling" scenario~\cite{Page:2006ly} in the
absence of the DU reactions, for $M<M_{c}^{\rm DU}$, the most
efficient cooling processes in a non-superfluid part of a NS are
the modified Urca (MU) processes $N+n\to N+p+l+\bar{\nu}_l$ and
$N+p+l\to N+n+{\nu}_l$, for $N=n,p$ and $l=e,\mu^-$. The soft-mode
bulk viscosity due to the MU processes was studied
in~\cite{Sawyer:1989dp,Haensel:2001mw}, where the FOPE model of
\cite{Friman:1978zq} was used for the description of the $NN$
interaction. The efficiency of so-calculated MU processes ($\sim
10^6$ times less than that would be for the DU processes) is
insufficient to stabilize the $r$-modes. Thus, the problem of the
$r$-mode instability becomes severe for the stars with masses
$M<M_{c}^{\rm DU}$.

In  case when the $NN$ interaction amplitude is mainly controlled
by the soft-pion exchange, the MU matrix element should be
replaced by the matrix element for the medium modified Urca (MMU)
processes, which for densities $n\gsim n_0$ are mainly determined
by the diagrams shown in Fig.~\ref{fig:MMU-react}. Here the bold
wavy line depicts the in-medium pion. The hatched vertices are
dressed by $NN$ correlations. Calculations~\cite{Voskre-cool} show
that the dominant contributions to the MMU rate come from the
first two diagrams in Fig.~\ref{fig:MMU-react}, whereas the third
diagram, which would be a naive generalization of the
corresponding MU (FOPE) contribution, gives only a small
correction for $n \gsim n_0$. We use the same parameterization of
the $NN$ interaction, degree of the pion softening, and
contributions to the matrix element of the neutrino reactions as
in~\cite{Blaschke:2013vma,BGV}. With these estimates, the
bulk viscosity owing to the MMU reactions increases by three to
four orders of magnitude for densities $n\sim 3n_0$, in accordance
with the corresponding increase of the neutrino emissivity of the
MMU processes studied in~\cite{Blaschke:2013vma,BGV}. For
densities $n>n_c^{\pi}$, there may appear  a pion condensate, as
the consequence of  enhancement of the pion softening with a
density increase.  We consider two possibilities: pion condensate
appears at $n_c^{\pi}$, and it does not appear provided the pion
softening saturates at higher densities. The contribution to the
bulk viscosity owing to reactions on a charged pion condensate, --- pion Urca (PU)
reaction, $n+\pi^-_{\rm c}\to n+l+\bar{\nu}_l$, --- is included in the
case when the condensate appears, at the critical density  taken
to be  $n_{c}^{\pi}=3n_0$, as in~\cite{Blaschke:2013vma,BGV}.
%%%%%%%%%%%%%%%%%%%%%%%%%%%

\begin{figure}
\includegraphics[width=8.cm]{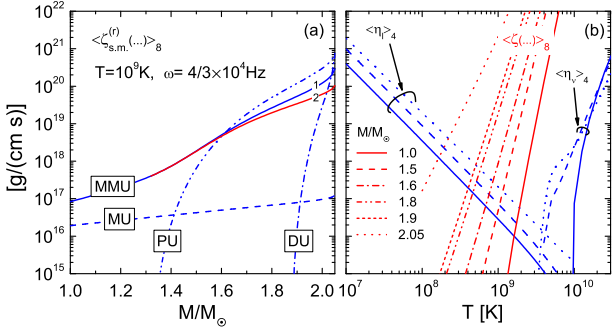}
\caption{Panel (a): Partial contributions to the bulk viscosity
averaged  over the star density profile, $\langle\zeta^{\rm
(r)}_{\rm s.m.} (1+0.86r^2/R^2)\rangle_8$, from DU, MU, MMU, and
PU reactions as functions of the NS mass. Line 1 shows the
contribution of the MMU reaction in absence of the pion
condensate and line 2, in its presence.
Nucleon pairing is not included.
%%%%%%%%
Panel (b): The profile-averaged  lepton $\langle\eta_l\rangle_4$
and neutrino $\langle\eta_{\nu}\rangle_4$ shear viscosity terms,
and the averaged total bulk viscosity $\langle\zeta
(1+0.86r^2/R^2)\rangle_8$, being calculated for the neutrino MMU
reactions, plotted as functions of the temperature for several NS
masses. Nucleon pairing is included. Calculations are done for
$\omega=(4/3)\cdot10^4$~Hz. \label{fig:zeta} }
\end{figure}

To illustrate the role of the pion softening, partial contributions to the bulk viscosity from different reactions averaged over the star density profile,
$\langle\zeta^{\rm (r)}_{\rm s.m.}[1+0.86 r^2/R^2] \rangle_8$, are
presented in Fig.~\ref{fig:zeta}a for MU, MMU, PU, and DU
reactions as functions of the NS mass for the temperature
$T=10^9$~K, but without account for pairing, and for the $r$-mode frequency $\om_4=4/3$. We see that for
$1\lsim M<1.6M_{\odot}$ the main contribution to $\langle
\zeta_{{\rm s.m.}}\rangle_8$ comes from the MMU processes. For
$M>1.6 M_{\odot}$ the PU process yields the dominant contribution.
Note that $\langle\zeta^{\rm (DU)}_{\rm s.m.}\rangle_8 <
\langle\zeta^{\rm (PU)}_{\rm s.m.}\rangle_8$ even for the heaviest
NS, since $M_c^{\rm DU}$ is rather close to the maximum mass for
our EoS. The MU and MMU contributions depend on the temperature as
$\propto T^6$ and the PU and DU ones as $\propto T^4$. All
contributions depend on the $r$-mode frequency as $\propto
\om^{-2}$. Pairing effects do not change the relative balance of partial contributions to $\zeta_{\rm s.m.}$.

The energy can be dissipated not only via non-equi\-librium
soft-mode processes but also by the neutrino radiation. Hence,
there is another source of the bulk viscosity, the radiative
viscosity. The latter term was studied only recently
in~\cite{Sa'd:2009vx} for the MU and DU reactions only. We
calculated contributions to the radiative bulk viscosity from MMU,
PU and DU reactions. Smaller contributions come from the processes
on weak neutral currents, such as nucleon bremsstrahlung reactions
and nucleon pair breaking-formation processes in superfluid
regions. We found that the radiative viscosity from MMU, PU and
in-medium nucleon bremsstrahlung processes demonstrates a strong
density dependence.

The shear viscosity contains several important contributions.
These are the terms from the lepton $\eta_{l}$ and neutron-neutron
scattering, $\eta_{n}$, and the neutrino contribution, $\eta_\nu$,
existing for temperatures, when neutrinos are trapped in the star
interior. The phonon contribution to the shear viscosity is found
to be  small for our choice of the pairing gaps.

The lepton shear viscosity term computed following
Ref.~\cite{Shternin:2008es}  proves to be by an order of magnitude
smaller than the term, computed previously
in~\cite{FlowersItoh,Cuttler} and exploited in many papers
studying $r$-modes. In spite of this suppression, the lepton
contribution to the shear viscosity proves to be larger than the
neutron one calculated with the free $NN$ cross-sections. In the
regions with the proton pairing the lepton term is
enhanced~\cite{Shternin:2008es}. We incorporate the in-medium
effects into $NN$ interaction amplitude and re-calculate  the
nucleon shear viscosity $\eta_{n}$, which turns to be still
smaller than the lepton one.

Neutrinos are trapped in the NS for temperatures above the opacity
temperature $T_{\rm opac}$ and, hence, can contribute to the shear
viscosity. Such a contribution was not considered yet. Usually,
$T_{\rm opac}$ is defined as the temperature, above which the
neutrino mean free path  $\lambda_\nu(\bar{n},T)$ at some averaged
nucleon density $\bar n$ becomes less than the NS radius,
$\lambda_\nu(\bar{n},T_{\rm opac})=R$. Evaluation of the opacity
temperature with the MU processes~\cite{Friman:1978zq} yields
$T_{\rm opac}\simeq 22\cdot 10^9$~K and this quantity weakly
depends on the averaged density. The account for the MMU reactions
leads to a decrease of $T_{\rm
opac}$~\cite{Voskre-cool,Migdal:1990vm}. The above definition of
$T_{\rm opac}$ does not take into account a strong density
dependence of the MMU reactions, which makes possible that  a
dense interior  is yet opaque  and a broad outer part is already
transparent for neutrinos. The radius of the region opaque for
neutrinos is determined from the condition $\lambda_\nu(n(r_{\rm
opac}),T)=R-r_{\rm opac}$. The opacity temperature can now be
defined as the temperature, at which $r_{\rm opac}=0$. The latter
temperature proves to be significantly smaller than the opacity
temperature introduced previously. We evaluate the neutrino shear
viscosity term with account for the pion-softening effects in the
neutrino mean free-path. Yet, for densities $n>n_c^{\pi}$ the pion
condensation processes contribute to the neutrino shear viscosity
and for the heaviest stars we add the contribution from the DU
processes.\\
\indent In Fig.~\ref{fig:zeta}(b) we collect our results  for shear and
bulk viscosities averaged over the star density profiles, as they
enter in times $\tau_\eta$ and $\tau_\zeta$. The viscosities are
computed with account for the nucleon pairing and plotted as
functions of the temperature for various star masses. The main
contributions to the bulk viscosity come from the soft-mode and
radiative terms. The main contribution to the shear viscosity is
the lepton term for $T\lsim 3\cdot 10^9$~K. It decreases with a
temperature increase, and at $T\gsim 4\cdot 10^9$~K the neutrino
contribution starts dominating for middle-heavy and heavy neutron
stars, whereas for the light neutron stars (with a mass $M\sim
M_{\odot}$) it happens at $T \gsim 10^{10}$~K. The bulk viscosity
rises rapidly with growth of the temperature and the NS mass. At
temperatures above $\sim 2\cdot 10^9$ for all star masses the bulk
viscosity exceeds the shear viscosity. For the  heaviest stars the
averaged bulk viscosity is dominated by the PU and DU reactions,
it exceeds the shear viscosity term already at $T>(2-5)\cdot
10^8$~K.\\
\begin{figure}\centerline{
\includegraphics[width=7.0cm]{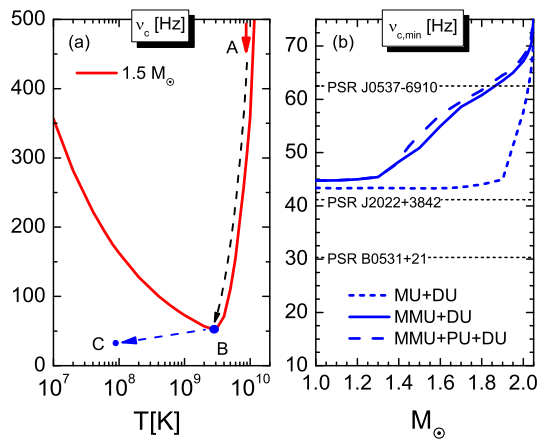}
}
\caption{ Panel (a): Critical rotation frequency for the $r$-mode
instability as a function of the star  temperature for the star
mass $1.5\,M_\odot$. Panel (b): The minimal critical frequency of
the star stable rotation [point B in panel (a)]  as a function of
the star mass, for different sets of reactions contributing to the
bulk viscosity. Dotted lines show rotation frequencies of three
most rapidly rotating pulsars. \label{fig:eta-nu} }
\end{figure}
\indent
The equation $\tau^{-1}(\nu_{\rm c}) =0$ determines the critical rotation frequency of a star. For frequencies larger than $\nu_c$ the $r$-modes amplitude, $a$, rapidly
grows unless non-linear effects stabilize it at some
maximal amplitude $a_{\rm max}$, for smaller frequencies the
$r$-modes are not excited. The $r$-modes stability line on the
temperature-frequency plane $(T,\nu_{\rm c})$ with
viscosities calculated within the nuclear medi\-um cooling scenario
is shown in Fig.~\ref{fig:eta-nu}(a) for the star with the mass
$1.5\,M_\odot$. The newly born pulsar enters the plane in the
upper-right corner (point A). The heat transport to the NS surface
which includes the charged-lepton  and nucleon thermal
conductivity, neutrino opacity, reheating by the $r$-modes etc.,
delays the cooling process.
The star trajectory on the $(T,\nu_c)$ plane closely follows
the critical line $\nu_c (T)$ between points A and B, provided the
typical cooling time is larger than the spin-down time $\sim 100{\rm s}/a_{\rm max}^2\nu_3^6$, where $\nu_3=\nu/10^3$Hz. For $a_{\rm max}\sim 1 $ the point B
with the frequency $\nu_{\rm c, min}$ corresponds to the
minimum of $\nu_{\rm c}$ as a function of $T$.
Cooling further down, the star
enters into the $r$-mode stable region and its rotation frequency
decreases at a much larger time scale determined by the
magnetic dipole radiation [path from B to C in
Fig~\ref{fig:eta-nu}(a)].
So, to explain the observed fast rotation frequency of the pulsar within this scenario the value of $\nu_{\rm c, min}$ must be larger than its current rotation frequency.

The quantity $\nu_{\rm c, min}$ is shown in
Fig.~\ref{fig:eta-nu}(b) as a function of the pulsar mass for
different sets of neutrino reactions contributing to the soft-mode
bulk viscosity. Within the minimal cooling  scenario when
in-medium effects are not included [short dashed line in
Fig.~\ref{fig:eta-nu}(b)], $\nu_{\rm c, min}$ exceeds $\nu_{\rm
max}^{\rm young}$ only for the masses $M>2.03M_{\odot}$ (when DU
reaction is already efficient), i.e. very close to the maximum
mass, $2.05M_{\odot}$. However, if initially the star passes
through the instability region, the developed $r$-modes blow off
some part of the star matter. So, its final mass (in point B) can
hardly be very close to the maximum mass. Alternatively, the
experimental value of the frequency of the pulsar PSR J0537-6910
could be explained within the minimal cooling scenario, if one
exploited EoS that allows for a lower DU threshold density.
Reference~\cite{Alford:2012yn} involving the data on $\dot{\nu}$ and $\nu$ presented arguments that within their analysis the curve AB in Fig.~\ref{fig:eta-nu}(a) is shifted to the left for the relevant values of the amplitude $a_{\rm sat}$, that could also allow to explain stability of PSR J0537-6910 within the minimal cooling paradigm.
Within the nuclear medium cooling scenario, the DU processes are not
needed to explain the stability of PSR J0537-6910. We explain it
for $M>1.80M_{\odot}$, if the  MMU and PU processes are included
or for $M>1.84 M_{\odot}$, if  pion condensate is not formed.

Concluding, we re-calculated contributions to the shear and bulk
viscosities for all dissipation processes considered previously
and we included new processes. We computed critical frequencies
for the $r$-mode stability of rotating neutron stars. The
in-medium polarization effects such as the Coulomb screening in
the electron-muon plasma and the softening of pionic degrees of
freedom with an increase of the nucleon density are incorporated.
The latter effect is the same as used in the nuclear-medium
cooling scenario applied successfully to the description of
neutron star cooling~\cite{Blaschke:2013vma,BGV}. The stability of
the most rapidly rotating young pulsar PSR J0537-6910 is explained
by the efficient MMU reactions provided its mass $M\geq
1.8M_{\odot}$. Presence of the DU reactions is not required.
Within the minimal cooling scenario the data on PSR J0537-6910 can
be explained only, if EoS allows for the efficient DU processes.
Finally, we should notice that there exist old pulsars in the LMXB
sources, which have frequencies much higher than  PSR J0537-6910.
It is commonly believed that these pulsars are accelerated by
accretion from the companion stars. They should be treated
separately.

The work was supported by Grants VEGA 1/0457/12 and APVV-0050-11 and by ``NewCompStar'', COST Action MP1304. We thank Dr. Kai Schwenzer for discussions.

%%%%%%%%%%%%%%%%%%%%%%%%%%%%%%%%%%%%%%%%%%%%%%%%%%%%%%%%%%%%%%%%%%

\end{document}